\documentclass[letterpaper,12pt]{article}

\usepackage{amsmath,amssymb}
\usepackage{epsfig}

\usepackage{axodraw}
\usepackage{color}
\usepackage{pstricks}

\begin{document}

\baselineskip=18pt

\newcommand{\eps}{\epsilon}
\newcommand{\pslash}{\!\not\! p}
\newcommand{\qslash}{\!\not\! q}
\newcommand{\ppslash}{\!\not\! p^{\,\prime}}
\newcommand{\m}{\widetilde m_u}
\newcommand{\M}{\widetilde M_u}
\newcommand{\md}{\widetilde m_d}
\newcommand{\Md}{\widetilde M_d}
\newcommand{\Real}{\mathsf{Re}}
\newcommand{\e}[1]{\varepsilon_#1}

\newcommand{\mchm}{MCHM$_5$}
\newcommand{\MCHM}{\mbox{MCHM$_{10}$}}

%%%%%%%%%%
%%%%%%%%%%    Title page
%%%%%%%%%%

\thispagestyle{empty}
\vspace{20pt}
\font\cmss=cmss10 \font\cmsss=cmss10 at 7pt

\begin{flushright}
UAB-FT-619 \\
ROMA1-1445/2006
\end{flushright}

\hfill
\vspace{20pt}

\begin{center}

{\Large \bf
Light custodians in natural composite Higgs models \\[0.2cm] 
}

\end{center}

\vspace{15pt}
\begin{center}
{\large  Roberto Contino$\, ^a$, Leandro Da Rold$\, ^b$ and
 Alex Pomarol$\, ^{c}$} 

\vspace{20pt}
$^{a}$\textit{Dipartimento di Fisica, Universit\`a di Roma ``La Sapienza'' and INFN \\
P.le A. Moro 2, I-00185 Roma, Italy}
\\[0.2cm]
$^{b}$\textit{Instituto de F\'isica, Universidade de S\~{a}o Paulo, \\
R. do Mat\~{a}o Travessa R, 187, 05508-900 SP, Brazil}
\\[0.2cm]
$^{c}$\textit{IFAE, Universitat Aut{\`o}noma de Barcelona,
08193 Bellaterra, Barcelona}

\end{center}

\vspace{20pt}
\begin{center}
\textbf{Abstract}
\end{center}
\vspace{5pt} {\small \noindent
We present a class of composite Higgs models arising from a warped extra dimension that
can satisfy all the electroweak precision tests 
in a significant portion of their parameter space.
A custodial symmetry plays a crucial role in  keeping the largest corrections to 
the electroweak observables below their experimental limits.
In these models the heaviness of the top quark 
is not only  essential to trigger the electroweak symmetry breaking, 
but it also implies that the lowest top resonance and its custodial partners, 
the custodians, are significantly lighter than the other resonances.
These custodians are the 
trademark of these scenarios.
They are exotic colored fermions of electromagnetic charges
$5/3$,  $2/3$ and $-1/3$, with masses predicted roughly in the range $500-1500$ GeV.
We discuss their production and detection at the LHC.
}

\vfill\eject
\noindent

%%%%%%%%%%
%%%%%%%%%%    Main Text
%%%%%%%%%%

\section{Introduction}

Theories of warped extra dimensions, with their holographic
interpretation in terms of 4D strongly coupled  field theories
\cite{Maldacena:1997re,Arkani-Hamed:2000ds},
have recently given a new twist to the idea of Higgs 
compositeness~\cite{Contino:2003ve,Agashe:2004rs}.
Calculability is one   of the  main virtues of this new class of models. 
Differently from the old approach~\cite{GK}, 
physical quantities of central interest can be computed in a perturbative expansion.
This opened up the possibility of building
predictive and realistic theories of electroweak symmetry 
breaking (EWSB).

The minimal composite Higgs model (MCHM) of Ref.~\cite{Agashe:2004rs}
is extremely simple to define based only on symmetry considerations.
It consists in a 5D theory on AdS spacetime compactified 
by two boundaries, respectively called infrared (IR) and ultraviolet  
(UV) boundary~\cite{Randall:1999ee}.~\footnote{Although these 
boundaries act as sharp cutoffs of 
the extra dimension, they can be considered as 
an effective description of some smoother
configuration that  can arise in a more  fundamental (string) theory.}
An SO(5)$\times$U(1)$_X\times$SU(3)$_c$ gauge symmetry in the bulk is     
broken down to SO(4)$\times$U(1)$_X\times$SU(3)$_c$   
on the IR boundary, (with SO(4)$\sim$SU(2)$_L\times$SU(2)$_R$), delivering  four 
pseudo-Goldstone bosons that transform as a \textbf{4} of SO(4) and
are identified with the Higgs doublet.
On the UV boundary the bulk symmetry is reduced to 
the Standard Model (SM) gauge group
$G_{\rm SM}=$SU(2)$_L\times $U(1)$_Y\times$SU(3)$_c$,
where hypercharge is defined as $Y=X+T_3^R$.
Once the SO(5) bulk representations in which the SM fermions are embedded and 
their boundary conditions are chosen, the model is completely determined. 
One can write down the  most general Lagrangian compatible with
the above symmetries, compute the one-loop Higgs potential and
determine the region of the parameter space with 
the correct EWSB and SM fermion masses.

In Ref.~\cite{Agashe:2004rs}
the SM fermions were embedded in spinorial representations of SO(5).
This choice leads  generically to large corrections to the $Zb_L\bar b_L$ coupling,
with the result that a sizable portion of the parameter space of the model ($\sim 95\%$) is
ruled out~\cite{Agashe:2005dk}.
However, it was recently realized~\cite{Agashe:2006at} 
that the $Zb_L\bar b_L$ constraint is strongly relaxed
if the fermions are embedded in fundamental ({\bf 5}) or antisymmetric ({\bf 10})
representations of SO(5), 
with $b_L$  belonging to a ${\bf (2,2)}$ of SU(2)$_L\times$SU(2)$_R$,
and the boundary symmetry SO(4) is enlarged to O(4).
In this case
a subgroup of the custodial symmetry O(3)$\subset$O(4)
protects the  $Zb_L\bar b_L$ coupling from receiving corrections.

In this paper we investigate the predictions of the MCHM with fermions in  
{\bf 5}'s or {\bf 10}'s  of SO(5) and the IR symmetry enlarged to O(4). 
We will determine the region of the parameter space with successful EWSB, 
and study the constraints imposed by the electroweak precision tests (EWPT).
The most relevant constraint comes from the Peskin-Takeuchi $S$ parameter, 
which excludes $\sim 50-75\%$ of the parameter space 
of the model.~\footnote{This must be compared with  the most popular supersymmetric models 
where  experimental constraints have already excluded large portions 
of the parameter space
($\sim 99\%$ in the case of universal soft masses~\cite{Giudice:2006sn}).} 
A sizable portion of the latter is still allowed, and awaits to be explored at the LHC.
An important prediction of the model is that the heaviness of the top quark
implies that the lowest top Kaluza-Klein (KK) resonance and 
its O(4)-custodial partners, the ``custodians'', are significantly lighter than the other KK resonances.
The custodians are color triplets and transform as $\bf 2_{7/6}$ of SU(2)$_L \times$U(1)$_Y$ 
when the SM fermions are embedded in ${\bf 5}$'s of SO(5),
and as 
${\bf 2_{7/6}\oplus 3_{2/3} \oplus 1_{5/3} \oplus 1_{-1/3}}$ in the case of 
${\bf 10}$'s of SO(5).
They have electromagnetic 
charges $Q_{\rm em}={ 5/3, 2/3, -1/3}$ and their masses are predicted 
roughly in the range $500-1500$ GeV.
The excitations of the SM gauge bosons are always heavier,
with masses around $2-3$ TeV.
The Higgs mass is predicted in the range $m_\text{Higgs} \simeq 115-190$ GeV.
Its value is correlated with the mass of the custodians,
since top loops give the largest contribution to the Higgs potential,
and are thus responsible for triggering the EWSB.
Being at the  reach of the LHC, the custodians
 offer the best  signature  for  distinguishing 5D composite Higgs 
from other scenarios of EWSB.  We will discuss 
their most important production mechanisms and 
decay channels.

Other models of EWSB in which $b_L$ is embedded in a $\bf {(2,2)}$ 
of SU(2)$_L \times$SU(2)$_R$ to protect $Zb\bar b$
have been proposed in Refs.~\cite{Carena:2006bn,Cacciapaglia:2006gp}.

\section{Higgs potential and EWSB}

At the tree level, the  massless spectrum of the bosonic sector of the MCHM
consists of the SM  gauge bosons, plus
four real scalar fields that
correspond to the SO(5)/SO(4) degrees of freedom of the fifth component of the 5D gauge field.
The presence of these scalars is dictated by the symmetry breaking pattern 
of the model: they are pseudo-Goldstone bosons and have 
the right quantum numbers to be identified with the SM Higgs, $h^a$ ($a=1,2,3,4$).
In addition to the massless sector, the theory also contains an infinite tower
of massive resonances: the KK states.
We can integrate out all the massive states 
and obtain an effective low-energy Lagrangian for the massless modes.
We do this by following the holographic approach of Ref.~\cite{Agashe:2004rs}.
The form of the effective Lagrangian for the gauge bosons
is completely determined by the symmetries of the model.
It can be found in Ref.~\cite{Agashe:2004rs}, and it will not be 
repeated here.~\footnote{The only difference between 
the gauge sector of the model presented here and that
of Ref.~\cite{Agashe:2004rs} is the symmetry on the IR boundary.
Enlarging SO(4) to O(4) forbids the otherwise allowed IR-boundary kinetic term  
$\epsilon^{(ijkl)}F_{(ij)}^{\mu\nu} F_{(kl)\, \mu\nu}$, 
where $i,j,k,l$ are SO(4) indices.
Since this term was not included in Ref.~\cite{Agashe:2004rs}, 
we can use the results for the gauge sector presented there.}
We only report the following relations:
\begin{equation} 
v \equiv \epsilon f_\pi 
= f_\pi \sin \frac{\langle h\rangle}{f_\pi}=246\ {\rm GeV} \, , \qquad
f_\pi = \frac{1}{L_1} \frac{2}{\sqrt{g_5^2 k}}\, ,\qquad
m_\rho\simeq \frac{3\pi}{4L_1}\, ,
\label{mw}
\end{equation}
where $h\equiv \sqrt{(h^a)^2}$.
Here $L_1$ denotes the position of the IR boundary 
in conformal coordinates and 
sets the mass gap  ($1/L_1\sim$ TeV);
$g_5$ is the SO(5) gauge coupling in the bulk;  $1/k$ is the AdS$_5$ curvature radius; and 
$m_\rho$ is the mass of the lightest gauge boson KK.
Following Ref.~\cite{Agashe:2004rs}, we define 
\begin{equation} \label{colors}
\frac{1}{N} \equiv \frac{g_5^2 k}{16\pi^2}
\end{equation}
as our expansion parameter.
The fermionic sector of the model 
depends on our choice for the 5D bulk multiplets.  
We want to study the case in which the SM fermions
are embedded in fundamental ($\bf 5$) or antisymmetric ($\bf 10$) representations
of SO(5). To this aim, we consider two different choices of multiplets and boundary conditions.
In the first choice, which we will refer to as the \mchm, 
the bulk fermions  transform as $\bf 5$'s of SO(5) and are defined by Eq.~(\ref{reps}) 
of the Appendix. In the second choice, which we will denote as the \MCHM,  
the bulk fermions are defined by Eq.~(\ref{fstates}) and transform as $\bf 10$'s of SO(5).
For all the technical details, we refer the reader to the Appendix. 
Here it will suffice to say that in both cases 
the low-energy effective Lagrangian for the quarks can be written,
in momentum space and at the quadratic order, as:
\begin{equation}
\begin{split}
{\cal L}_{\rm eff}= 
&\bar q_L \pslash \left[ \Pi_0^q(p^2) 
 + \frac{s^2_h}{2} \left( \Pi_1^{q1}(p^2)\, \hat H^c \hat H^{c\dagger} 
 +  \Pi_1^{q2}(p^2)\, \hat H \hat H^\dagger \right) \right] q_L \\
& +\bar u_R \pslash \left( \Pi_0^u(p^2) + \frac{s^2_h}{2}\, \Pi_1^u (p^2)\right) u_R 
 +\bar d_R \pslash \left( \Pi_0^d (p^2)+ \frac{s^2_h}{2}\, \Pi_1^d (p^2)\right) d_R  \\
&+ \frac{s_hc_h}{\sqrt{2}} M_1^u (p^2)\,\bar q_L \hat H^c u_R 
 + \frac{s_hc_h}{\sqrt{2}} M_1^d (p^2)\,\bar q_L \hat H d_R + h.c. \,  .
\end{split}
\label{effla}
\end{equation}
Here $c_h\equiv\cos (h/f_\pi)$, $s_h\equiv\sin (h/f_\pi)$, and
\begin{equation}
\hat H =\frac{1}{h} \begin{bmatrix}   h^1 - i  h^2 \\ 
  h^3 - i h^4 \end{bmatrix}\, , \quad
\hat H^c =\frac{1}{h} \begin{bmatrix} -( h^3 + i h^4) \\  
  h^1 + i h^2 \end{bmatrix}\, . \end{equation}
The form factors $\Pi^i_{0,1}$ and $M^i_{1}$ in Eq.~(\ref{effla}) can be 
computed in terms of 5D propagators using the 
holographic approach of Ref.~\cite{Agashe:2004rs}.
Their explicit form is given in the Appendix.
From Eq.~(\ref{effla}) one can derive the SM up and down fermion masses, $m_{u,d}$.
A reasonably good approximation to the exact expressions can be obtained 
by setting $p^2=0$ in the form factors, the error being of order $(m_{u,d}L_1)$.
We obtain
\begin{equation}
m_{u}\simeq \frac{s_hc_h}{\sqrt{2}}\frac{M_1^{u}(0)}{\sqrt{Z_{u_L} Z_{u_R}}}\, , \quad\quad
m_{d}\simeq \frac{s_hc_h}{\sqrt{2}}\frac{M_1^{d}(0)}{\sqrt{Z_{d_L} Z_{d_R}}}\, , 
\label{mf}
\end{equation}
where $Z_{u_L,d_L}=\Pi^q_0(0)+ (s^2_h/2) \, \Pi^{q1, q2}_1(0)$ and
$Z_{u_R,d_R}=\Pi^{u,d}_0(0)+ (s^2_h/2) \, \Pi^{u, d}_1(0)$.
An explicit version of Eq.~(\ref{mf})
in terms of the 5D input parameters 
is given in Eqs.~(\ref{muanalytic5}) and  (\ref{muanalytic10}) of the Appendix,
respectively for the \mchm\ and the \MCHM.

At the tree level the Higgs field is an exact Goldstone boson, and as such it has no potential.
At the one-loop level, the virtual exchange of the SM fields transmits the explicit breaking
of SO(5) and generates a potential for $h$.
The largest contribution comes from $t_L$, $b_L$ and $t_R$, and from the SU(2)$_L$ gauge bosons,
since these are the fields that are most strongly coupled to the Higgs.
They give
\begin{equation}
\begin{split}
V(h) =& \frac{9}{2} \int\! \frac{d^4p}{(2\pi)^4} \log\left(\Pi_0 +\frac{s^2_h}{4}\, \Pi_1 \right)
       -2 N_c \int\!\frac{d^4p}{(2\pi)^4}\bigg\{
       \log \left( \Pi_0^q + \frac{s^2_h}{2}\, \Pi_1^{q2} \right)\\[0.2cm]
       &+ \log\left[
         p^2 \left( \Pi_0^u + \frac{s^2_h}{2}\, \Pi_1^u \right) 
         \left( \Pi_0^q + \frac{s^2_h}{2}\, \Pi_1^{q1} \right)  
         -\frac{s^2_h c^2_h}{2} (M_1^{u})^2 \right] \bigg\}\, .
\end{split}
\label{pot}
\end{equation}
Here, and from now on, the fermionic form factors $\Pi_0^q$, $\Pi_1^{q1,q2}$, $\Pi^u_{0,1}$,
$M_1^u$ stand for those of the 3rd quark family, $q_L = (t_L,b_L)$ and $t_R$.
The gauge form factors $\Pi_{0,1}$ can be found in Ref.~\cite{Agashe:2004rs}.
Since the $\Pi_1^i$'s and $M_1^u$  drop exponentially  for $p L_1\gg 1$, 
the logarithms in Eq.~(\ref{pot}) can be expanded and the 
potential is well approximated by
\begin{equation}
V(h)\simeq \alpha\ s^2_h-\beta\ s^2_hc^2_h\, ,
\label{potential}
\end{equation} 
where $\alpha$ and $\beta$ are integral functions of the form factors.
In particular, $\alpha$ receives contributions
from loops of the gauge fields and of $q_L$ or $t_R$ alone.
We will denote these contributions respectively by $\alpha_{\rm gauge}$, $\alpha_L$ and $\alpha_R$.
On the other hand, $\beta$ receives contributions from loops where both $t_L$
and $t_R$ propagate.
For $\alpha< \beta$ and  $\beta\geqslant 0$ we have that the electroweak symmetry is broken: 
$\epsilon\not=0$. 
If $\beta>|\alpha|$, the minimum of the potential is at
\begin{equation}
s_h=\epsilon=\sqrt{\frac{\beta-\alpha}{2\beta}}\, ,
\label{vev}
\end{equation}
while for $\beta<|\alpha|$ the minimum corresponds to $c_h=0$, and the 
EWSB is maximal: $\epsilon=1$.
In this latter case the fermion masses vanish
-- see Eq.~(\ref{mf}) --
due to an accidental chiral symmetry that is restored
in the limit $\eps\to 1$.
The model is thus realistic only for $0<\epsilon<1$,  i.e. $\beta>|\alpha|$. 
The coefficients $\alpha$ and $\beta$ 
can be computed in terms of the relevant 5D parameters
\begin{equation}
N\, ,\  c_q\, ,\ c_u\, ,\  \m \, ,\  \M \, ,
\label{parameters}
\end{equation}
where  $c_q$, $c_u$ are  the bulk masses
(in units of  $k$) 
of the 5D multiplets $\xi_q$, $\xi_u$ that 
contain  the SM $q_L$ and $t_R$, 
and $\m$, $\M$
are  mass terms  localized on the IR boundary
that mix $\xi_q$ with  $\xi_u$ (see Appendix).
The scale $L_1$ has been traded for $v$. 
The  five parameters  of Eq.~(\ref{parameters})
are not completely determined by the present experimental data. 
There are only two constraints coming  from fixing the top quark mass to its experimental
value, $m_t^{pole} = 173$ GeV, and by requiring $0<\eps<1$.

Let us outline the viable region of this five-dimensional space of parameters.
The large  Yukawa coupling of the top quark is reproduced 
by having the wave functions of the $t_L$ and $t_R$ zero modes 
peaked towards the IR boundary, where the Higgs lives.
This constrains the 5D bulk masses to lie in the interval $|c_{q,u}|<1/2$, see 
Eqs.~(\ref{muanalytic5}) and (\ref{muanalytic10}).
In the 4D dual interpretation, this corresponds to say that the elementary fields
$t_L$ and $t_R$ couple to relevant operators ${\cal O}$ of the strongly coupled conformal
field theory (CFT), with conformal dimension $3/2<\text{Dim}[{\cal O}]<5/2$ 
(see~\cite{Contino:2004vy}). 
Since the operators ${\cal O}$ have the quantum numbers to excite
the fermionic composite resonances, the elementary top will have a sizable mixing 
with these massive states. 
The stronger the mixing, 
the larger will be the degree of compositeness of the physical top quark.
The requirement $|c_{q,u}|<1/2$ is also necessary in order to have EWSB.
In this region the top quark contribution to the Higgs potential dominates 
over the gauge one, which would otherwise align the vacuum in an 
(SU(2)$_L\times$U(1)$_Y$)-preserving direction (since  $\alpha_{\rm gauge}$ is always positive).
The conditions $\alpha<0$ and $\beta>0$ can then be easily satisfied 
thanks to  the top contribution.  
In other words, the EWSB, in our model, is a direct consequence of the heaviness of the top.

\begin{figure}[th]
\begin{minipage}{\linewidth}
\epsfig{file=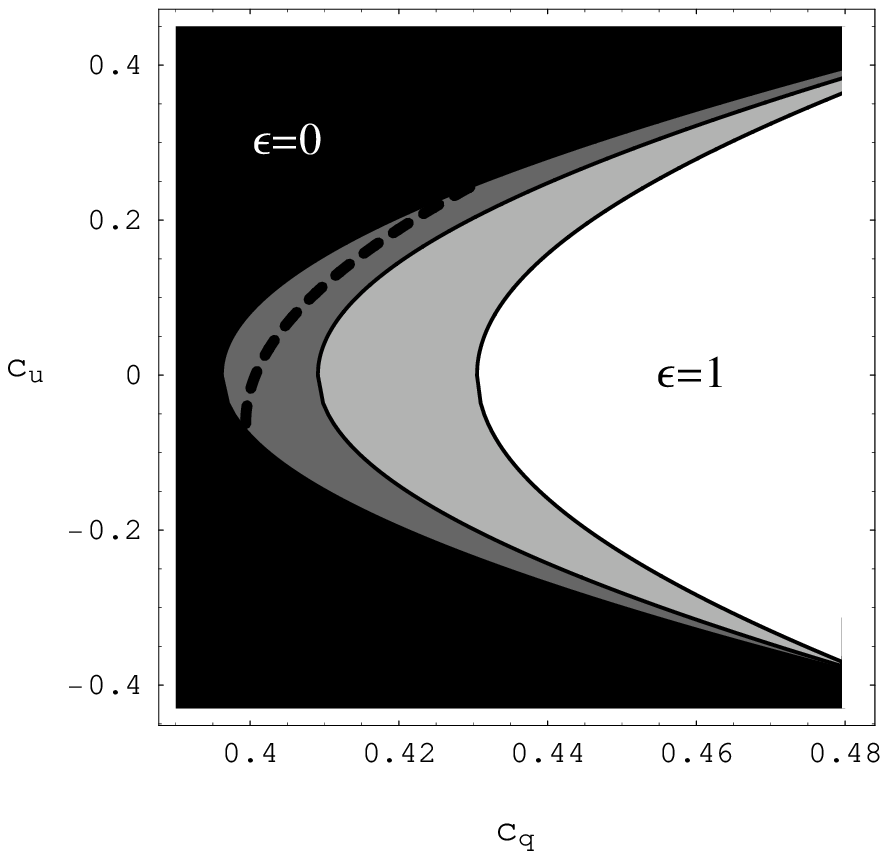,width=0.498\linewidth}
\quad
\epsfig{file=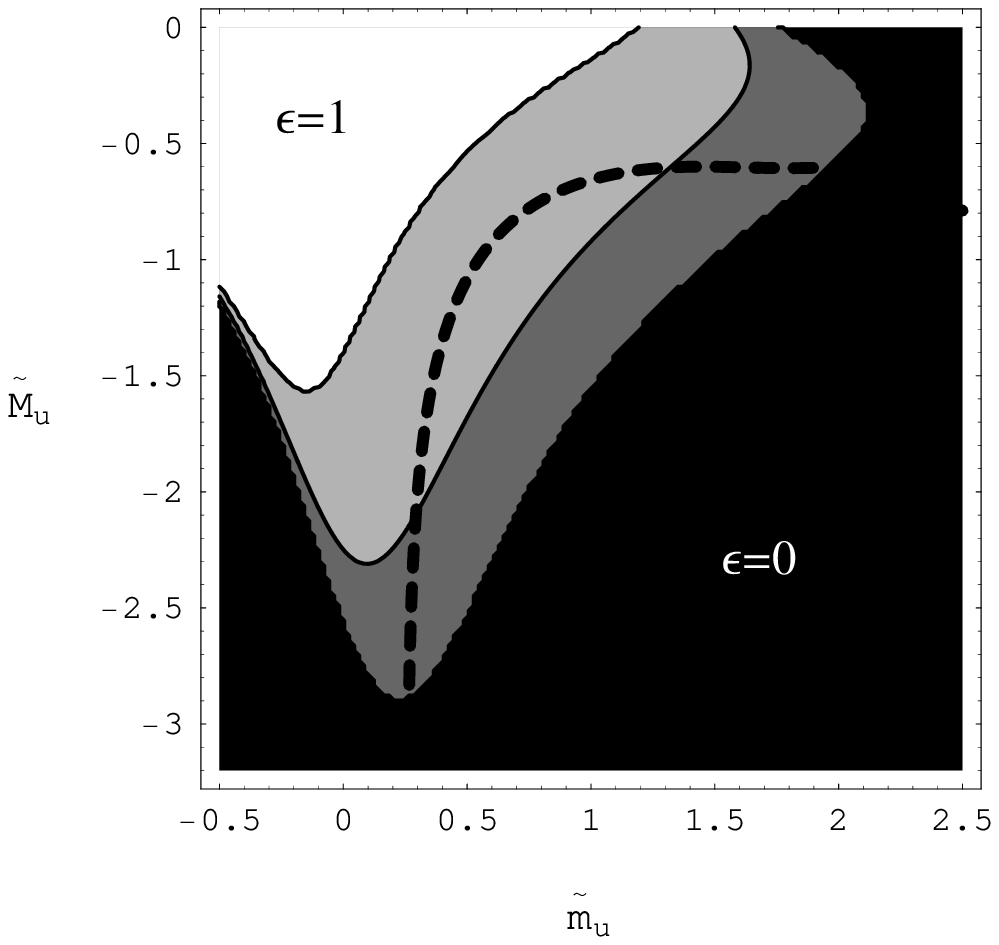,width=0.498\linewidth}  
\end{minipage}
\caption{\it \small
Contour plots of $\eps$ in the plane $(c_q,c_u)$ 
with $\m=1$, $\M = -2$, $N=8$ 
for the \MCHM\ (left plot),
and in the plane $(\m,\M)$ with $c_q=0.35$, $c_u=0.45$, $N=8$ for the \mchm\ (right plot).
The two gray  areas   correspond to  the  region with
EWSB and non-zero fermion masses, $0 < \eps < 1$.
The  lighter gray area  is excluded when  
the bound $S\lesssim 0.3$ is imposed.
The dashed black line represents the curve with 
$m_t^{\overline{\text{MS}}}(2\,\text{TeV})= 150\,$GeV,
equivalent to $m_t^\text{pole} = 173\,$GeV.
}
\label{fig:contours}
\end{figure}

To illustrate this point, we show in Figure~\ref{fig:contours} the
contour plots of $\eps$ 
in the planes $(c_q,c_u)$ and $(\m,\M)$ respectively for the \MCHM\ and the \mchm.
The region with no EWSB ($\eps=0$) is depicted in black,
and the dashed black curve corresponds to $m_{t}^{pole} = 173$ GeV.
In each plot, we have set $N=8$ and kept the remaining 5D parameters fixed.
The condition $\beta>|\alpha|$, \textit{i.e.}  
$0 < \eps < 1$, further selects a smaller region of the planes $(c_q,c_u)$ and $(\m,\M)$. 
A naive estimate shows that
$|\alpha_{L,R}|$ is parametrically larger than $\beta$ by a factor $(1/4-c^2_{u,q})$.
A reduction in  $\alpha$, however,  can be obtained in the region where 
$\alpha_L\simeq  -\alpha_R$.  
As already pointed out in Ref.~\cite{Agashe:2004rs}, this is possible since 
$\alpha_L$ and $\alpha_R$ have generally opposite sign.
In the case of the \MCHM, the region with smaller $\alpha$ are the two gray areas
with the ``boomerang'' shape
shown in the left plot of Fig.~\ref{fig:contours}, plus a specular region 
under $c_q \to -c_q$ which is not showed. These two solutions correspond to
$q_L$ almost elementary ($c_q \simeq +1/2$) or almost composite ($c_q \simeq -1/2$).
We found that the case of the \mchm\ is analogous, but with the role of $c_q$ and $c_u$
interchanged: the two possible solutions are for 
$t_R$ almost elementary ($c_u \simeq -1/2$) or almost composite ($c_u \simeq +1/2$).

A second circumstance in which $\beta>|\alpha|$
is when $\m \simeq -1/\M$.
As one can directly check, by using their expressions in terms of 5D propagators, 
the fermionic form factors $\Pi_1^{q1}$, $\Pi_1^u$, and consequently 
$\alpha_{L,R}$,~\footnote{Notice that $\Pi_1^{q2} = 2 \Pi_1^{q1}$ in the \MCHM, while
in the \mchm\ $\Pi_1^{q2}$ is always suppressed and its contribution to
$\alpha_L$ can be neglected, see Appendix.}
identically vanish for $\m = -1/\M$ (both in the \mchm\ and in the \MCHM).
Therefore, for $\m \simeq  -1/\M$  one can have $\beta>|\alpha|$.
This is shown for   the \mchm\   in Fig.~\ref{fig:contours}, right plot, 
and a similar result holds for the \MCHM.
Even though one can reach the $0<\eps <1$ region by moving along the plane $(\m,\M)$ 
for almost any choice of $c_q$, $c_u$,
the additional constraint of having the top quark mass equal to its experimental value 
(the black dashed line in the plots of Fig.~\ref{fig:contours})
disfavors in most of the cases solutions with both $q_L$ and $t_R$ elementary 
(that is, with $c_q\simeq 1/2$ and $c_u\simeq -1/2$). 
This is especially true for the \MCHM, since the formula for the top mass has an 
extra suppression factor $1/\sqrt{2}$ compared to the \mchm, see Eqs.~(\ref{muanalytic5}) 
and (\ref{muanalytic10}) of the Appendix.

Our investigation of the structure of the Higgs potential has then revealed that
there are specific regions of the parameter space in which 
the electroweak symmetry is broken and the SM quarks get a mass.
Part of these regions, however, is excluded by the precision tests.
How large is this portion gives
us a measure of the ``degree of tuning'' required in our model.
This is the subject of the next section.

\section{Electroweak precision tests}
\label{sec:EWPT}

There are two types of corrections to the electroweak observables 
that   any composite Higgs model must address, since they are usually sizable:
Non-universal  corrections to the $Zb\bar b$ coupling, and universal corrections 
to the gauge boson self-energies.  
The results of Ref.~\cite{Agashe:2006at} show that 
for both our choices of fermionic 5D representations, Eqs.~(\ref{reps}) and (\ref{fstates}), 
non-universal  corrections to  $Zb\bar b$ are small, 
due to the custodial O(3) symmetry of the bulk and IR boundary.
Therefore, we need to consider only universal effects, which can be 
parametrized in terms of four quantities: $S$, $T$, $W$ and $Y$~\cite{Barbieri:2004qk}.
The last two parameters are  suppressed by a factor $\sim (g^2 N/16\pi^2)$ compared 
to $S$ and $T$, and can be neglected~\cite{Agashe:2004rs}.
The parameter $T$
is zero at tree level due to  the custodial symmetry.
Loop effects can be estimated to be small  ($T\lesssim 0.3$), and explicit calculations
in similar 5D models  confirm this 
expectations~\cite{Agashe:2005dk,Carena:2006bn}.~\footnote{\label{foot}
In Ref.~\cite{Carena:2006bn} the SM fermions were also embedded in 
the ${\bf 5}$ and ${\bf 10}$  representations of SO(5), but with different
boundary conditions from ours.
This implies that, for example,  while
in the \mchm\ there is one SU(2)$_L$-doublet KK state
with hypercharge $Y=7/6$ that becomes light in the limit of $t_R$ composite, 
in the model of Ref.~\cite{Carena:2006bn} this 
happens in the limit of $t_R$ mostly elementary.
}
We defer a full calculation of the $T$ parameter in the \mchm\ and \MCHM\ to a future
study.

The Peskin-Takeuchi $S$ parameter gives the most robust and model-independent constraint.
Neglecting a small correction from boundary kinetics terms, one has~\cite{Agashe:2004rs}:
\begin{equation}
S = \frac{3}{8} \frac{N}{\pi} \epsilon^2\, .
\end{equation}
The $99\%$ CL experimental bound $S\lesssim 0.3$~\cite{Barbieri:2004qk}
\footnote{
In order to fully saturate the bound $S\lesssim 0.3$, a positive $T$ 
of the same size is required. 
The results of Refs.~\cite{Agashe:2005dk,Carena:2006bn} suggest that this
could be possible   in certain regions of the parameter space. 
Otherwise, the $99\%$ CL experimental bound  on $S$ becomes stronger: 
$S\lesssim 0.2\, (0.1)$ for $T\lesssim 0.1\, (0)$.}
then translates into
\begin{equation}
\epsilon^2\,\lesssim \frac{1}{4}\left(\frac{10}{N}\right)\, .
\label{cons}
\end{equation}
For $N=5$ ($N=10$) this rules out the values $1/2\lesssim\epsilon^2<1$ ($1/4\lesssim\epsilon^2<1$),
which we naively expect to correspond to $\sim 1/2$ ($\sim 3/4$) of the  region
with EWSB and non-zero fermion masses (the region $0<\epsilon^2<1$).
The exact numerical results -- see Fig.~\ref{fig:contours} for a case with $N=8$ -- reasonably agree
with this rough estimate.
This means that there is still a large portion of the parameter space which is not ruled out
by the constraint from $S$, and no large fine tuning is hence required.
Notice that  smaller values of $N$ imply larger fractions of allowed parameter space,
although $N$ cannot be too small if we want to remain in the perturbative regime.

\section{Spectrum of fermionic resonances  and the Higgs mass}

An important  prediction of our model is that the requirement of  
a large top  mass always forces some of the
fermionic KK states  to be lighter than their gauge counterpart (a similar 
property   is found in  the model of Ref.~\cite{Panico:2006em}).
The reason is the following. 
The embedding of $t_L$ and $t_R$ into SO(5) bulk multiplets
implies that some of their SO(5) partners have
$(\pm,\mp)$ boundary conditions, an assignment that is necessary 
to avoid extra massless states (see Eqs.~(\ref{reps}) and (\ref{fstates})).
Consider for example the case in which the left-handed chirality of the
5D field is $(+,-)$, (hence the right-handed one is $(-,+)$): for values
of the 5D mass $c_{i= u, q}> 1/2$, the lightest KK mode,  
denoted  by $q^*$,  has its left-handed chirality
exponentially peaked on the UV boundary, while the right-handed one is
localized on the IR boundary. This implies that the mass of $q^*$
is exponentially suppressed. On the other hand, for $c_i<-1/2$
both chiralities are localized on the IR boundary and the mass of $q^*$ is of the same
order of that of the other KKs: $m_{q^*}\simeq m_\rho$.
In the intermediate region $-1/2<c_i<1/2$, the one 
in which the large mass of the top can be reproduced,
one finds that $m_{q^*}$
is well interpolated by~\cite{Agashe:2004rs,Agashe:2005dk}
\begin{equation}
m_{q^*} \simeq \frac{\kappa}{L_1}\, \sqrt{\frac{1}{2}-c_i}\, ,
\label{mqs}
\end{equation}
where $\kappa \sim 2$ is a numerical coefficient with a mild dependence on the values of the
boundary masses.
This means that $m_{q^*}$ is still parametrically smaller than $m_\rho$ by a factor 
$\sqrt{{1}/{2}-c_i}$.
Analogous results hold for left-handed 5D fermions  with $(+,+)$ boundary condition
if they mix with  additional  4D  fermions localized on the IR boundary.
In the case of right-handed 5D fields  with
$(+,-)$ boundary condition
the same argument above also applies if $c_i\rightarrow -c_i$.
Eq.~(\ref{mqs})   has a clear interpretation 
in the 4D dual description of the theory where
the left- and right-handed chiralities of the lightest massive state 
correspond respectively to an elementary and a composite state~\cite{Contino:2004vy}.
For $-1/2<c_i<1/2$, it can be shown (see also the Appendix) that 
the coupling of the elementary state to the CFT flows
to a fixed point value proportional to $\sqrt{-\gamma}= \sqrt{{1}/{2}-c_i}$, 
where $\gamma$ is the anomalous dimension of the CFT operator to which the
elementary field couples. 
Naive dimensional analysis then immediately gives Eq.~(\ref{mqs}).

We will concentrate on the region $-1/2<c_u<1/2$ with $c_q$ slightly smaller than
$1/2$ ($t_L$ mostly elementary).
From the argument above and 
by inspecting Eqs.~(\ref{reps}) and (\ref{fstates}),
one can  deduce that 
the light KK modes are all the SO(5)  partners of $t_R$ inside $\xi_u$.
In the case of the \mchm, $\xi_u$ transforms as a ${\bf 5}$ of SO(5)
and the partners of $t_R$ form a $\bf (2,2)$ of SU(2)$_L \times$SU(2)$_R$,
equivalent to two SU(2)$_L$ doublets of hypercharge $Y=1/6$ and $Y=7/6$. 
The first is the lightest resonance with the quantum numbers of $t_L$, while
the second is its O(4)-custodial companion.
In the case of the \MCHM, $\xi_u$ transforms as a ${\bf 10}$ of SO(5)
and the light partners of $t_R$ also include its own O(4) custodians,
a ${\bf 3_{2/3} \oplus 1_{5/3} \oplus 1_{-1/3}}$ of SU(2)$_L \times$U(1)$_Y$.

\begin{figure}
\centering
\epsfig{file=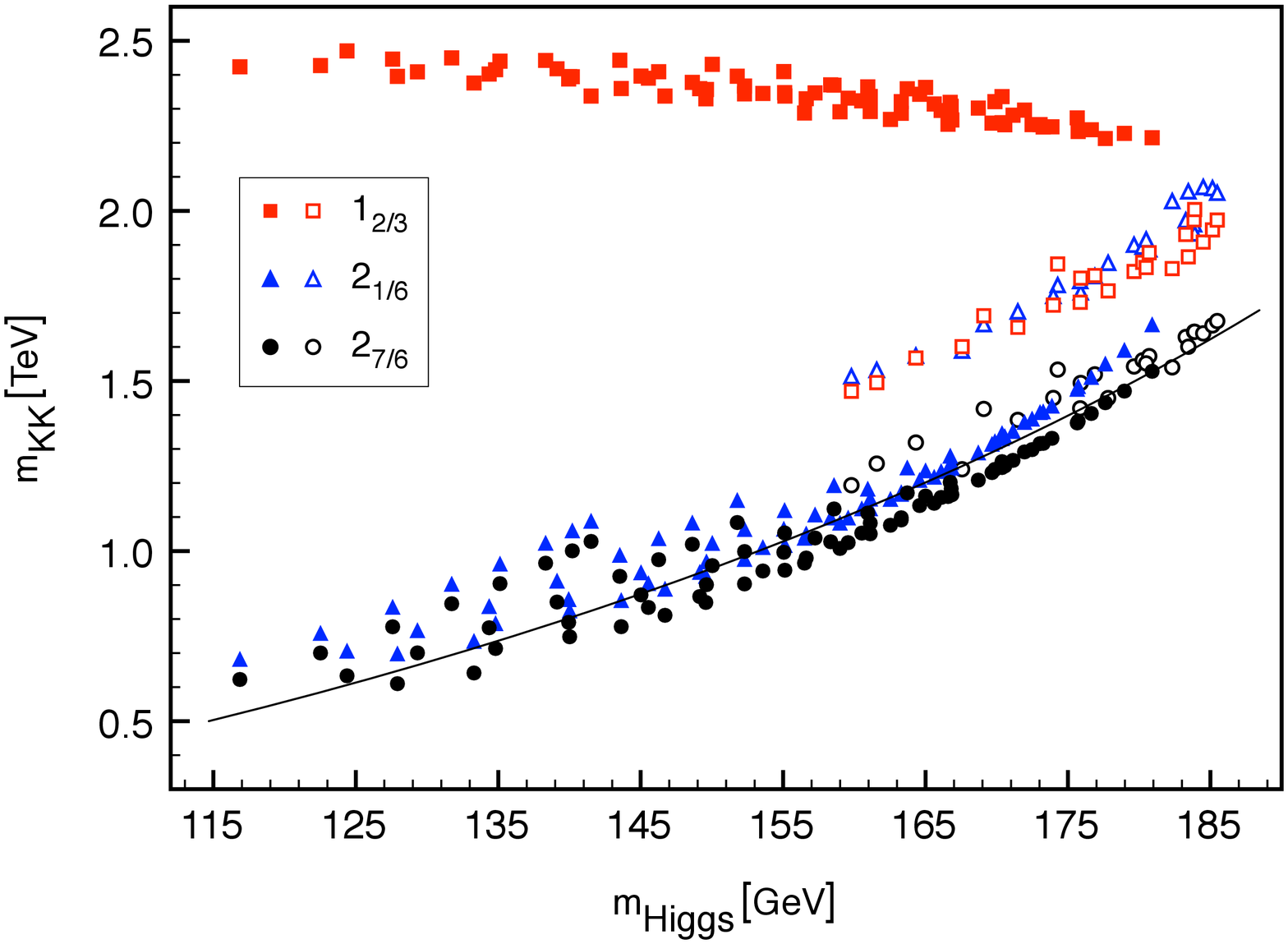,width=0.70\linewidth} \\[0.5cm]
\epsfig{file=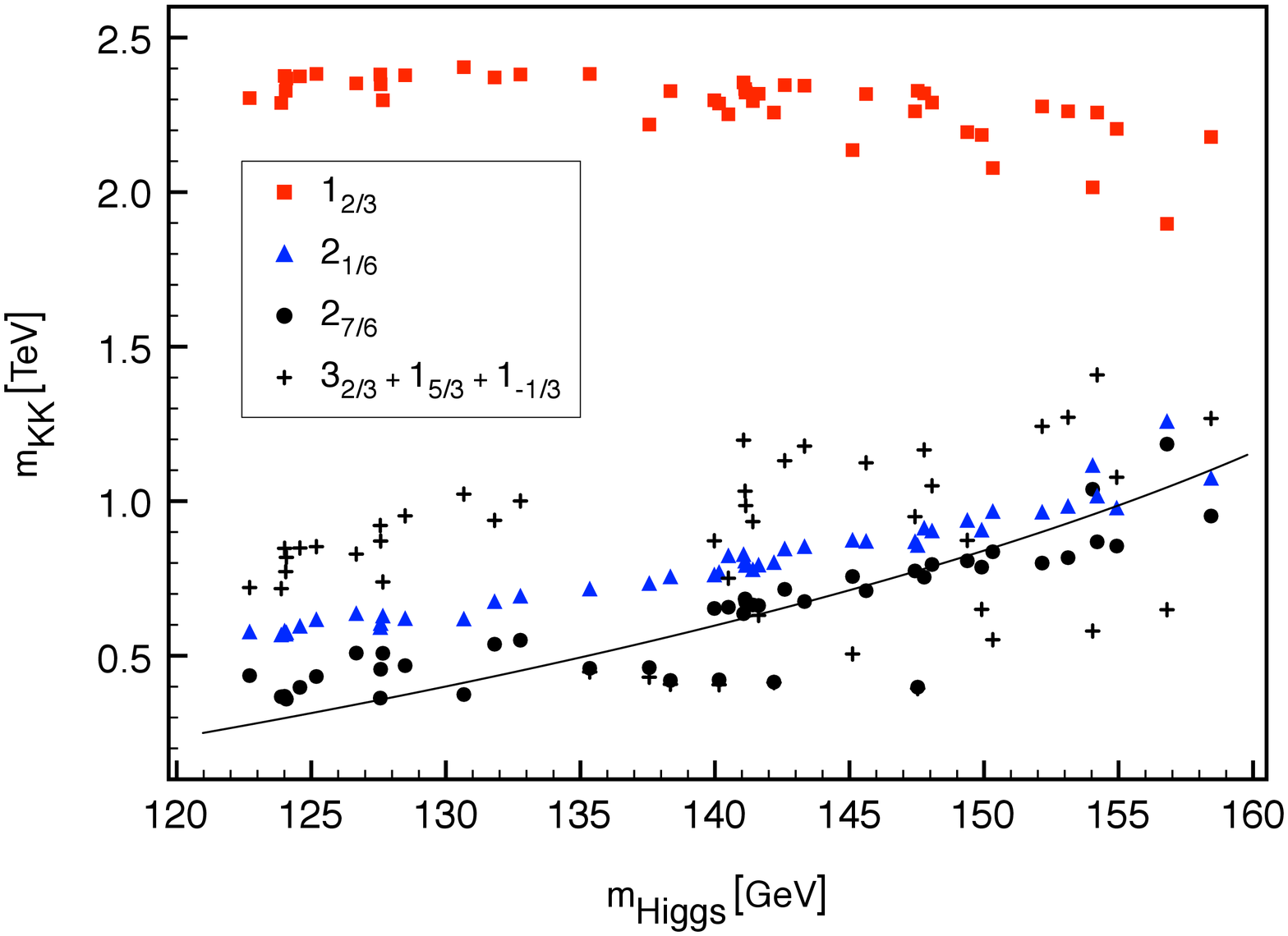,width=0.70\linewidth}
\caption{\it
Masses of the lightest colored KK fermions
in the \mchm\ (upper plot), and in the \MCHM\ (lower plot).
Different symbols denote KKs with different 
quantum numbers under SU(2)$_L\times$U(1)$_Y$, as specified in the plots. 
Both plots are for $\eps = 0.5$, $N=8$.
In the upper one we have varied
$0.28 < c_q < 0.38$, $0 < c_u < 0.41$, $0.32 < \m < 0.42$, $-3.5 < \M < -2.2$
(filled points), or 
$0.2 < c_q < 0.35$, $-0.25 < c_u < -0.42$, $-1.3 < \m < 0.2$, $0.1 < \M < 2.3$
(empty points).
In the lower plot we have varied
$0.36 < c_q < 0.45$, $0 < c_u < 0.38$, $0.8 < \m < 3$, $-3 < \M < -0.3$.
The black continuous line is the fit to the mass of the lightest resonance
according to  Eqs.~(\ref{cutoff}) and (\ref{lacut}).
}
\label{fig:mKK}
\end{figure}

Figure~\ref{fig:mKK} shows the spectrum of the lowest  fermionic KK states
in the \mchm\  (upper plot), and in the \MCHM\  (lower plot).
The light states are those predicted. Their mass 
is around $500-1500\,$GeV for $\eps=0.5$ and $N=8$,  much smaller than that of
the lightest gauge KK, $m_\rho = 2.6\,$TeV, and of other fermionic excitations.
The custodian ${\bf 2_{7/6}}$ turns out to be  the lightest among all the light fermionic
resonances and therefore the most accessible.

There is an alternative way to understand why in this type of models
one expects light   colored fermionic resonances.
From Eq.~(\ref{potential}), we have that the Higgs mass is given by 
\begin{equation}
m_h^2\simeq \frac{8\beta}{f^2_\pi}\, s^2_hc^2_h\, ,
\label{mh2}
\end{equation}
where 
\begin{equation}
\beta\simeq N_c\int \frac{d^4p}{(2\pi)^4}\frac{F(p)}{p^2}\, ,
 \qquad F(p)\equiv \frac{(M^u_1)^2}%
           {\left(\Pi^q_0+ (s^2/2)\,\Pi^{q1}_1\right)\left(\Pi^{u}_0+ (s^2/2)\,\Pi^{u}_1\right)}\, .
\end{equation}
Using Eq.~(\ref{mf}), we have $m_t^2=F(0)s^2_hc^2_h/2$ and hence
\begin{equation}
m_h^2\simeq \frac{N_c}{\pi^2}\frac{m^2_t}{v^2}\, \epsilon^2\Lambda^2\, ,
\label{cutoff}
\end{equation}
where we have defined 
\begin{equation}
\Lambda^2 \equiv 2\int^\infty_0 \!\!\! dp\; p\, \frac{F(p)}{F(0)}\, .
\end{equation}
Eq.~(\ref{cutoff}) 
shows the relation between the Higgs mass and $\Lambda$, which is, 
roughly speaking, the scale at which the momentum integral
is cut off.
\enlargethispage{0.5cm}
On general grounds, one would expect this cutoff scale
to be of the order of the mass of the lowest fermionic resonance:
\footnote{Using Eq.~(\ref{vev})  we can rewrite 
Eq.~(\ref{cutoff})  as 
\begin{equation*}
m^2_h\simeq  \frac{N_c}{\pi^2}\frac{m^2_t}{2v^2}\Lambda^2-\frac{4c^2_h\alpha}{f^2_\pi}\, .
\label{SMqd}
\end{equation*}
The first term is the formula for the Higgs mass one obtains in the SM by defining
$\Lambda^2/2 \equiv \int\! dp\; p $ in the top loop.
The degree of cancellation between the first and  second term 
gives a measure of the degree of ``tuning'' needed in our model.
This  exactly corresponds to $\epsilon^2$. 
}
\begin{equation}
m_{q^*}\simeq \Lambda \simeq 900\ \text{GeV} \left(\frac{m_h}{150\ \text{GeV}}\right)
\left(\frac{0.5}{\eps}\right)\, ,
\label{mqs2}
\end{equation} 
where in the last equality Eq.~(\ref{cutoff}) has been used.
Eq.~(\ref{mqs2})  shows that  in composite Higgs models with a light Higgs
and no tuning ($\eps\sim 1$) 
 colored resonances are expected to be not heavier than  $\sim$ 1 TeV.
In our model, the relation between the Higgs mass and the mass of the lowest fermionic KK turns out
to be more complicated than that of Eq.~(\ref{mqs2}).
We find that the points of Fig.~\ref{fig:mKK} are better reproduced by a relation of the form
\begin{equation}
\Lambda^2 = a_1\, m^2_{{q^*}}+a_2\, m_{q^*} M + a_3\, M^2\, ,
\label{lacut}
\end{equation}
where $a_i$ are numerical coefficients, 
$M \equiv m_\rho$  parametrizes the mass of the heavier resonances and
by $m_{q^*}$ we denote the mass of the 
KK weak doublet with hypercharge $Y=7/6$
(the lightest among the fermionic KKs in Fig.~\ref{fig:mKK}).
This means that in our model 
the integral $\int^\infty_0 dp\, p\,[F(p)/F(0)]$ is  
not completely cut off at $p\sim m_{{q^*}}$,
and that other (heavier) resonances also play a role.
A fit to the points of Fig.~\ref{fig:mKK} gives:
 $a_{i=1,2,3}=(-0.10,0.35,0.007)$ for  the \mchm\ (upper plot) and 
$a_i=(-0.14,0.24,0.06)$ for the \MCHM\  (lower plot).
The dispersion of the points around the fitted curve (shown in each plot) 
can be explained as follows.
In  Fig.~\ref{fig:mKK} we have fixed  $N =8$, $\eps = 0.5$ and $m_t^{pole} = 173$ GeV,
which leaves two of the five parameters of Eq.~(\ref{parameters}) free to vary.
If  $c_u$ is traded for $m_{q^*}$,
we are left with one free parameter, for example,  $c_q$. 
The coefficients $a_i$ of Eq.~(\ref{lacut}) will thus depend on $c_q$.
To generate the points
in Fig.~\ref{fig:mKK}  we have scanned over the values $0.2 < c_q < 0.38$ (upper plot)
and $0.36 < c_q < 0.45$ (lower plot)
and therefore the fitted $a_i$ given above should  be considered
as average values over these intervals of $c_q$.

\section{Production and detection of the lightest  fermionic resonances at the LHC}

The most promising way to discover  these models
is by detecting their
lowest fermionic KKs at the LHC.
In particular, detecting the custodian  with electric charge $5/3$,
$q^*_{5/3}$,
that arises from the  $\mathbf{2}_{7/6}$ multiplet
of SU(2)$_L\times$U(1)$_Y$, would be the 
smoking-gun signature of the model.  
This exotic state  
is  a direct consequence of the custodial symmetry required to
forbid large corrections to $Z b\bar b$.
For not-too-large values of its mass $m_{q^*_{5/3}}$, roughly below 1 TeV, this new particle will 
be mostly produced in pairs, via QCD interactions,
\begin{equation}
q\bar q , \,  gg \to q^*_{5/3}\, \bar q^*_{5/3} \, ,
\label{pro}
\end{equation}
with a cross section completely determined in terms of $m_{q^*_{5/3}}$ 
(see for example~\cite{Han:2003wu,Azuelos:2004dm}).
Once produced, $q^*_{5/3}$  will decay  to a (longitudinally polarized)
$W^+$ plus a $t_R$, with a coupling of order $4\pi/\sqrt{N}\, \sqrt{(c_{u}+1/2)}$.
Decays to SM light quarks will be strongly suppressed, with a coupling of order
of the square root of their Yukawa couplings.
In general, colored resonances will mostly decay to tops and bottoms,
since these are the SM quarks more strongly coupled to them, as the
result of the large top mass.
The process of Eq.~(\ref{pro}) then leads to a final state of four $W$'s and two $b$-jets:
\begin{equation}
q^*_{5/3}\, \bar q^*_{5/3} \to W^+ t \, W^- \bar t \to W^+ W^+ b \, W^- W^- \bar b \, .
\end{equation}
The same final state also comes
from pair production of KKs with charge $-1/3$.
A way to discriminate between the two cases consists in 
reconstructing the electric charge of the resonance.
For example, one could look for events with two highly-energetic
leptons of the same sign, coming from the leptonic decay of two of the four $W$'s,
plus at least six jets, two of which tagged as $b$-jets.
Demanding that the invariant mass of the system of the two hadronically-decaying
$W$'s plus one $b$-jet equals $m_{q^*_{5/3}}$ then identifies the process and gives
evidence for the charge $5/3$ of the resonance.
Furthermore, indirect evidence in favor of ${q^*_{5/3}}$  would
come from the non-observation of the decays to $Z b$, $H b$
that  are allowed for resonances of charge $-1/3$.

For increasing   values of $m_{q^*_{5/3}}$ the cross section for pair production quickly drops,
and single production might become more important.
The relevant process is $tW$ fusion~\cite{Willenbrock:1986cr}, 
where a longitudinal $W$ radiated from one proton
scatters off a top coming from the other proton.
The analogous process initiated by a bottom quark, $bW$ fusion, has been studied in detail
in the literature
and shown to be an efficient way to singly produce
heavy excitations of the top quark~\cite{Han:2003wu,Perelstein:2003wd,Azuelos:2004dm}.
To prove that the same conclusion also applies to the case of $tW$ fusion a dedicated analysis
is required. The main uncertainty and challenge comes from the small top quark content of the proton,
which however can be compensated by the large coupling involved, especially in the case
of $t_R$ largely composite~\cite{Agashe:2004rs}.

Besides  $q^*_{5/3}$, the other components of the $\mathbf{2}_{1/6}$ and 
$\mathbf{2}_{7/6}$ multiplets of 
SU(2)$_L\times$U(1)$_Y$ are also predicted to be light in both our models, see 
Fig.~\ref{fig:mKK}.
In the specific case of the \MCHM\ there are also other states transforming as 
$\mathbf{3}_{2/3}$, 
$\mathbf{1}_{5/3}$, $\mathbf{1}_{-1/3}$.
These multiplets  contain 
 states  with $Q_{\rm em}=5/3$ whose
 phenomenology will be similar to that of the $q^*_{5/3}$ described above.
 The states of electric charge $2/3$ or 
$-1/3$ will also be produced in pairs via QCD interactions 
or singly via $bW$ or $tW$ fusion.
They will decay to a SM top or bottom quark plus a longitudinally polarized $W$ or
$Z$, or a Higgs.
When kinematically allowed, a heavier resonance will also decay to a lighter KK accompanied  
with  a  $W_{\rm long}$,  $Z_{\rm long}$ or $h$.
Decay chains could lead to extremely characteristic final states. For example, in the 
\MCHM\ the KK with charge $2/3$ from $\mathbf{3}_{2/3}$ is predicted
to be generally heavier than $q^*_{5/3}$, see Fig.~\ref{fig:mKK}.
If pair produced, it can decay to $q^*_{5/3}$ leading to a spectacular six $W$'s plus 
two $b$-jets final state:
\begin{equation}
q^*_{2/3}\, \bar q^*_{2/3} \to W^- q^*_{5/3}\, W^+ \bar q^*_{5/3} \to 
W^- W^+ W^+ b \, W^+ W^- W^- \bar b \, .
\end{equation}
To fully explore the phenomenology of the fermionic resonances in our models
a detailed analysis is necessary.
One could for example adopt the simplifying strategy proposed in Ref.~\cite{twosite},
where a 4D effective theory has been introduced to consistently describe the SM fields
and the first KK excitations of a large class of warped models.
Existing studies in the literature have focussed on the production and detection
of SU(2)$_L$ singlets of hypercharge $Y=2/3$~\cite{Azuelos:2004dm,singlet}, 
since this is a typical signature of Little Higgs theories. In our models, however, the singlet is not 
predicted to be light, except for specific regions of the parameter space.
In conclusion, our brief discussion shows that there are characteristic signatures
predicted by our models that will distinguish them from  other extensions of the SM.
While certainly challenging, these signals will be extremely spectacular, and will provide
an  indication of  a new strong dynamics responsible for 
electroweak symmetry breaking.

%%%%%%%%%%
%%%%%%%%%%    Acknowledgments
%%%%%%%%%%

\section*{Acknowledgments}

It is a pleasure to thank
Kaustubh Agashe for collaboration in the early stages of this project
and for many useful discussions and comments.
R.C. also thanks G.~Servant for useful discussions.
The work of L.D. and A.P. was  partly supported   by the
FEDER  Research Project FPA2005-02211
and DURSI Research Project SGR2005-00916.
The work of L.D. was partly supported by FAPESP.
A.P. and R.C. thank the Galileo Galilei Institute for Theoretical Physics for 
hospitality and the INFN for partial support during the completion of this work.
A.P. also thanks the theory group at CERN for hospitality.

%%%%%%%%%%
%%%%%%%%%%    Appendix
%%%%%%%%%%

\appendix

\section*{Appendix}

Here we present in detail the fermion sector of our composite Higgs models.
We have assumed that:
(\textit{i}) each SM fermion is embedded in a different 5D field;
(\textit{ii}) All the three SM families have the same embedding.
Following these rules, we 
construct  what we think are  the minimal models
with fermions embedded in {\bf 5} or {\bf 10} representations
of SO(5).

\section{Defining the MCHM$_5$}

The quark sector of the \mchm\ is defined in terms of 5D bulk multiplets
transforming as fundamentals of SO(5).
Each SM generation is identified with the zero modes
of the 5D fields
\begin{equation} \label{reps}
\begin{alignedat}{2}
\xi_{q_1} &= \!\left[\!\!
 \begin{array}{ll}
   \mathbf{(2,2)^{q_1}_L} = \begin{bmatrix} q'_{1L}(-+) \\ q_{1L}(++) \end{bmatrix} &
   \mathbf{(2,2)^{q_1}_R} = \begin{bmatrix} q'_{1R}(+-) \\ q_{1R}(--) \end{bmatrix}  \\[0.5cm]
   \mathbf{(1,1)^{q_1}_L}(--)   & \mathbf{(1,1)^{q_1}_R}(++)   
 \end{array}  \!\!\right]\! , \;\;
\xi_u &= 
 \begin{bmatrix}
\mathbf{(2,2)^{u}_L}(+-) & \mathbf{(2,2)^{u}_R}(-+)   \\[0.15cm]
\mathbf{(1,1)^{u}_L}(-+)   & \mathbf{(1,1)^{u}_R}(+-)
 \end{bmatrix}  ,  \\[0.7cm]
\xi_{q_2} &= \!\left[\!\!
 \begin{array}{ll}
   \mathbf{(2,2)^{q_2}_L} = \begin{bmatrix} q_{2L}(++) \\ q'_{2L}(-+) \end{bmatrix} &
   \mathbf{(2,2)^{q_2}_R} = \begin{bmatrix} q_{2R}(--) \\ q'_{1R}(+-) \end{bmatrix}  \\[0.5cm]
   \mathbf{(1,1)^{q_2}_L}(--)   & \mathbf{(1,1)^{q_2}_R}(++)   
 \end{array}  \!\!\right]\! , \;\;
\xi_d &= 
 \begin{bmatrix}
\mathbf{(2,2)^{d}_L}(+-) & \mathbf{(2,2)^{d}_R}(-+)   \\[0.15cm]
\mathbf{(1,1)^{d}_L}(-+)   & \mathbf{(1,1)^{d}_R}(+-)
 \end{bmatrix}  ,
\end{alignedat}
\end{equation}
where $\xi_{q_1}$,  $\xi_{u}$ ($\xi_{q_2}$,  $\xi_{d}$) transform as \textbf{5}$_{2/3}$
(\textbf{5}$_{-1/3}$) of SO(5)$\times$U(1)$_X$.
A similar 5D embedding also works for the SM leptons, although with different
U(1)$_X$ charges.
Chiralities under the 4D Lorentz group have been denoted by $L$, $R$, and 
$(\pm,\pm)$ is a shorthand notation to denote Neumann $(+)$ or Dirichlet $(-)$
boundary conditions on the two boundaries.
We have grouped the fields of each multiplet $\xi_i$ in
representations of SO(4)$\sim$SU(2)$_L\times$SU(2)$_R$, and used the fact
that a fundamental of SO(5) decomposes as 
$\mathbf{5} = \mathbf{4} \oplus \mathbf{1}  = \mathbf{(2,2)} \oplus \mathbf{(1,1)}$.
Although each of $\xi_{q_1}$ and $\xi_{q_2}$ alone could account for the $q_L$ zero mode, 
we need them both to give mass to both the up and down SM quarks.
We thus identify the SM $q_L$ field with the zero mode of the linear combination 
$(q_{1L}+q_{2L})$, and get rid of the extra zero mode by introducing a localized 
right-handed field on the UV boundary that has a mass mixing with the orthogonal combination.
We denote by $c_i$, $i=q_1 , q_2, u , d$, the  bulk masses of each 5D field $\xi_i$
in units of  $k$.
Localized on the IR boundary, we consider the most general set of mass terms
invariant under O(4)$\times$U(1)$_X$:
\begin{equation}
 \m\,  \mathbf{ \overline{(2,2)}_L^{q_1}}  \mathbf{(2,2)_R^{u}}  +
 \M\,  \mathbf{ \overline{(1,1)}_R^{q_1}}  \mathbf{(1,1)_L^{u}} +
 \md\,  \mathbf{ \overline{(2,2)}_L^{q_2}}  \mathbf{(2,2)_R^{d}}  +
 \Md\,  \mathbf{ \overline{(1,1)}_R^{q_2}}  \mathbf{(1,1)_L^{d}} + h.c.
\end{equation}

\subsection{The holographic description}

The 5D field content of Eq.~(\ref{reps}) has a very simple holographic 
interpretation in terms of three elementary chiral fields, $q_L = (u_L, d_L)$, $u_R$ and $d_R$,
coupled to a CFT sector through composite operators $O_i$.~\footnote{We adopt 
a left-source holographic
description for the fields $\xi_{q_1}$ and $\xi_{q_2}$, and a 
right-source description for $\xi_{u}$ and $\xi_{d}$.
See Ref.~\cite{Contino:2004vy}. }
An important difference from the model of Ref.~\cite{Agashe:2004rs}, and also from
the \MCHM\ discussed in the next section, is that here the elementary $q_L$
couples to two different CFT operators $O_{1}$ and $O_{2}$ with couplings $\lambda_{1}$, $\lambda_2$.
This is the consequence of having two different bulk fields in Eq.~(\ref{reps}), 
$q_{1L}$ and $q_{2L}$, mixed on the UV boundary.  
The elementary fields $u_R$ and $d_R$, instead, couple to one operator
each, $O_u$ and $O_d$, with couplings $\lambda_u$ and $\lambda_d$. 
The bulk gauge symmetry of the 5D model maps into an
SO(5)$\times$U(1)$_X$ global symmetry of the CFT, and Eq.~(\ref{reps})
implies that $O_1$, $O_u$ transform as \textbf{5}$_{2/3}$, while
$O_2$, $O_d$ transform as \textbf{5}$_{-1/3}$. Schematically:
\begin{center}
\hspace*{-2cm}
  \begin{picture}(166,91) (135,-198)
    \Text(152,-150)[c]{$q_L$}
    \SetWidth{0.8}
    \Line(165,-142)(200,-132)
    \Line(165,-158)(200,-168)
    \Text(185,-118)[t]{$\lambda_1$}
    \Text(185,-184)[b]{$\lambda_2$}
    \Text(209.5,-132)[lc]{$O_{1}$}
    \Text(209.5,-171)[lc]{$O_{2}$}
    \Line(231,-130.5)(253,-130.5)
    \Line(231,-133)(253,-133)
    \Line(231,-170)(253,-170)
    \Line(231,-172.5)(253,-172.5)
    \Text(260,-133)[lc]{$O_u$}
    \Text(260,-172)[lc]{$O_d$}
    \Text(242.5,-125)[bc]{$H$} \Text(242.5,-180)[tc]{$H$}
    \Line(279,-132)(309,-132) \Line(279,-171.5)(309,-171.5)
    \Text(323,-132.5)[c]{$u_R$} \Text(323,-171.5)[c]{$d_R$}
    \Text(295,-126)[bc]{$\lambda_u$} \Text(295,-180)[tc]{$\lambda_d$}
  \end{picture}
\end{center}
The double line indicates that $(\bar O_{1} O_u)$ and $(\bar O_{2} O_d)$ have the correct 
quantum numbers to excite the Higgs and generate in this way the up and down Yukawa couplings of the 
4D low-energy theory. Large hierarchies among the Yukawas can be explained naturally as the
result of the RG evolution of the couplings $\lambda_i$ \cite{Agashe:2004rs}.
Notice that the CFT dynamics alone do not mix $O_1$ and $O_u$ with $O_2$ and $O_d$,
since they have different U(1)$_X$ quantum numbers. 
Nevertheless, both $O_1$ and $O_2$ can couple to the external source $q_L$, since this latter
coupling will only preserve the SU(2)$_L \times$U(1)$_Y$ elementary symmetry.
This suggests that a hierarchy in the up and down Yukawa couplings, like in the case of the top and
bottom quarks, can  follow from the RG evolution 
if $\lambda_u \gg \lambda_d$ at low-energy, as already 
pointed out for the  model of Ref.~\cite{Agashe:2004rs}, or alternatively
if $\lambda_1 \gg \lambda_2$.

Quite interestingly, it is simple to show that $\lambda_1$ can grow
much bigger (or smaller) than $\lambda_2$ in the infrared, 
even if both operators $O_1$ and $O_2$ are relevant.
The argument goes as follows.
The RG equations of the two couplings $\lambda_1$, $\lambda_2$ form a coupled system:
\begin{align}
%\begin{split}
p \frac{d}{dp} \lambda_1 (p) &= \gamma_1\, \lambda_1 + \frac{N}{16\pi^2} 
 \left( a_1\, \lambda_1^3 + a_{12}\, \lambda_1 \lambda_2^2 \right) + \,\dots \\[0.1cm]
\label{RG2}
p \frac{d}{dp} \lambda_2 (p) &= \gamma_2\, \lambda_2 + \frac{N}{16\pi^2} 
 \left( a_2\, \lambda_2^3 + a_{21}\, \lambda_2 \lambda_1^2 \right) + \,\dots
\end{align}
The dots stand for subleading terms in a $1/N$ expansion, 
where the number of CFT colors $N$ is defined by Eq.~(\ref{colors}).
The duality with the 5D theory implies that 
the coefficients $a_1$ and $a_2$ are both positive (see~\cite{Contino:2004vy}), 
and that the anomalous dimensions $\gamma_{1,2}$ are linear functions of the 5D bulk masses:
$\gamma_{1,2} =| c_{q_1 , q_2} + 1/2|-1$.
Furthermore, it is easy to show that the leading contribution to the
coefficients $a_{12}$ and $a_{21}$ comes from the wave function renormalization, which
in turn implies $a_{21} = a_1$, $a_{12} = a_2$ at leading order in $1/N$.
Let us then consider the case in which the operator with the smallest anomalous dimension,
say $O_1$, is relevant, that is $c_{q_1} < 1/2$, $c_{q_1} < c_{q_2}$
({\it i.e.} $\gamma_1 < 0$, $\gamma_1 < \gamma_2$).
Then $\lambda_1$ will grow faster than $\lambda_2$ in the infrared, and at some
energy $E_*$ it will reach a fixed-point value 
$\lambda_{1*} \simeq 4\pi/\sqrt{N}\; \sqrt{-\gamma_1/a_1}$.
Below that energy, one can set $\lambda_1 = \lambda_{1*}$ in the RG equation for $\lambda_2$,
Eq.~(\ref{RG2}), which becomes
\begin{equation}
p \frac{d}{dp} \lambda_2 (p) \simeq \left( \gamma_2 - \gamma_1 \right) \lambda_2 +\,\dots
\end{equation}
Since $(\gamma_2 - \gamma_1) > 0$ by assumption, this implies that $\lambda_2$ will
be suppressed at low energy,
\begin{equation}
\lambda_2(E) \sim \left( \frac{E}{E_*} \right)^{\gamma_2-\gamma_1} \, ,
\end{equation}
even if $\gamma_2 < 0$, that is, even if the operator $O_2$ is relevant.

The above argument shows that  the small ratio
$m_b/m_t$ follows naturally in the \mchm\ by having $c_{q_2} > c_{q_1}$, even
when $b_R$ is strongly coupled to the CFT sector.
The large top mass, on the other hand, requires $|c_{q_1}|, |c_u| < 1/2$
for the third generation quarks, i.e. the operators $O_1$ and $O_u$ need to be
both relevant. In our numerical analysis of the \mchm\ (hence in all the results
presented in the text), we have set $c_{q_2} = 0.4$, $c_{d} = -0.55$ for the third generation,
and varied $-1/2 < c_u < 1/2$, $-1/2 < c_{q_1} < c_{q_2}$.
Choosing $c_{q_1} < c_{q_2}$ also implies that the contribution
of $\xi_{q_2}$ to the Higgs potential will be suppressed and hence negligible
compared to that of $\xi_{q_1}$ (see also below).  This justifies the following identification:
\begin{equation}
\xi_q \equiv \xi_{q_1} \, , \qquad c_q \equiv c_{q_1}\, ,
\end{equation}
where by $\xi_q$ we mean the field responsible for the contribution of $q_L$ 
to the Higgs potential to which we referred in the text.

Following the method of Ref.~\cite{Agashe:2004rs}, one can derive the most general form 
of the holographic Lagrangian by introducing spurion fields and embedding the elementary sources 
in complete SO(5)$\times$U(1)$_X$ (chiral) multiplets. 
The fact that the elementary $q_L$ couples to two different CFT operators implies that
there are two different ways to embed it in a fundamental representation of SO(5),
namely as the $T^{3R}=+1/2$ or the $T^{3R}=-1/2$ component of the internal $\mathbf{(2,2)}$. 
More explicitly, grouping the entries of each \textbf{5}
in SU(2)$_L\times$SU(2)$_R$ representations:
\begin{equation}
\Psi_{1L} = 
 \begin{bmatrix} 
  \begin{pmatrix} q'_{1L} \\ q_L \end{pmatrix} \\[0.4cm] u'_L
 \end{bmatrix} \, , \quad
\Psi_{2L} =
 \begin{bmatrix} 
  \begin{pmatrix} q_{L} \\ q'_{2L} \end{pmatrix} \\[0.4cm] d^{\,\prime}_L
 \end{bmatrix} \, , \quad
\Psi_{uR} = 
 \begin{bmatrix} 
  \begin{pmatrix} q^u_{R} \\ q^{\prime\, u}_{R} \end{pmatrix} \\[0.4cm] u_R
 \end{bmatrix} \, , \quad
\Psi_{dR} = 
 \begin{bmatrix} 
  \begin{pmatrix} q^{\prime \, d}_{R} \\ q^{u}_{R} \end{pmatrix} \\[0.4cm] d_R
 \end{bmatrix} \, .
\end{equation}
The multiplets $\Psi_{1L}$, $\Psi_{uR}$ ($\Psi_{2L}$, $\Psi_{dR}$) have U(1)$_X$ charge
$+2/3$ ($-1/3$), and all components other than $q_L$, $u_R$ and $d_R$
are non-dynamical spurion fields.
Therefore, the most general (SO(5)$\times$U(1)$_X$)-invariant holographic Lagrangian, 
at the quadratic order and in momentum space, is
\begin{equation} \label{effla5}
\begin{split}
{\cal L}_{holo}^{(2)} = \!
 & \sum_{r=1,2} \bar\Psi_{rL}^i \pslash \left( \delta^{ij} \hat\Pi_0^{rL}(p)
  + \Sigma^i \Sigma^j \hat\Pi_1^{rL}(p) \right)\! \Psi_{rL}^j
  + \! \sum_{r=u,d} \bar\Psi_{rR}^i \pslash \left( \delta^{ij} \hat\Pi_0^{rR}(p)
  + \Sigma^i \Sigma^j \hat\Pi_1^{rR}(p) \right)\! \Psi_{rR}^j \\[0.2cm]
 &+ \bar\Psi_{1L}^i \!\left( \delta^{ij} \hat M_0^{1L}(p)
   +\Sigma^i \Sigma^j \hat M_1^{1L}(p) \right)\! \Psi_{uR}^j 
  + \bar\Psi_{2L}^i \!\left( \delta^{ij} \hat M_0^{2L}(p)
    +\Sigma^i \Sigma^j \hat M_1^{2L}(p) \right)\! \Psi_{dR}^j
 + h.c.
\end{split}
\end{equation}
Here $i,j= 1,\dots, 5$ are SO(5) indices, and $\Sigma$ is the non-linear
realization of the Higgs field~\cite{Agashe:2004rs}:
\begin{equation}
\Sigma = \frac{s_h}{h} \left( h^1,h^2,h^3,h^4, h \; \frac{c_h}{s_h} \right)\, .
\end{equation}
The form factors $\hat\Pi$, $\hat M$ can be computed using the holographic technique 
of Ref.~\cite{Agashe:2004rs}; the result is:
\begin{equation} \label{finalff5-1}
\begin{alignedat}{2}
\hat\Pi_0^{1L} &= \Pi_{q_L}(c_{q_1},c_u,\m)\, , &
\hat\Pi_1^{1L} &= \Pi_{Q_L}(c_{q_1},c_u,\M)-\Pi_{q_L}(c_{q_1},c_u,\m)\, ,
\\[0.15cm] 
\hat\Pi_0^{2L} &= \Pi_{q_L}(c_{q_2},c_d,\md)\, , &
\hat\Pi_1^{2L} &= \Pi_{Q_L}(c_{q_2},c_d,\Md)-\Pi_{q_L}(c_{q_2},c_d,\md)\, ,
\\[0.3cm]
\hat\Pi_{0,1}^{uR} &= \hat\Pi^{1L}_{0,1} (c_{q_1} \leftrightarrow c_u ; L \leftrightarrow R)\, , \qquad &
\hat\Pi_{0,1}^{dR} &= \hat\Pi^{2L}_{0,1} (c_{q_2} \leftrightarrow c_d ; L \leftrightarrow R)\, , 
\\[0.3cm]
\hat M_0^{1L} &= M_{q}(c_{q_1},c_u,\m)\, ,  &
\hat M_1^{1L} &= M_{Q}(c_{q_1},c_u,\M)-M_{q}(c_{q_1},c_u,\m)\, ,
\\[0.15cm] 
\hat M_0^{2L} &= M_{q}(c_{q_2},c_d,\md)\, , &
\hat M_1^{2L} &= M_{Q}(c_{q_2},c_d,\Md)-M_{q}(c_{q_2},c_d,\md)\, ,
\end{alignedat}
\end{equation}
where $\Pi_{q_L,Q_L}$ and $M_{q,Q}$ are the form factors defined in the Appendix of 
Ref.~\cite{Agashe:2004rs}. After setting all non-dynamical fields to zero, the Lagrangian
(\ref{effla5}) reduces to that of Eq.~(\ref{effla}) with
\begin{equation} \label{finalff5-2}
\begin{split}
\Pi_0^q &= \hat\Pi_0^{1L} + \hat\Pi_0^{2L} \, , \\
\Pi_0^u &= \hat\Pi_0^{uR} + \hat\Pi_1^{uR} \, , \\
\Pi_0^d &= \hat\Pi_0^{dR} + \hat\Pi_1^{dR} \, , 
\end{split} \qquad
\begin{split}
\Pi_1^{q_1}&= \hat\Pi_1^{1L}\, , \\[0.05cm]
\Pi_1^{q_2}&= \hat\Pi_1^{2L}\, , 
\end{split} \qquad 
\begin{split}
\Pi_1^{u}  &= -\frac{1}{2}\, \hat\Pi_1^{uR}\, , \\[0.15cm]
\Pi_1^{d}  &= -\frac{1}{2}\, \hat\Pi_1^{dR}\, , 
\end{split} \qquad
\begin{split}
M_1^u &= \hat M_1^{1L}\, , \\[0.05cm]
M_1^d &= \hat M_1^{2L}\, .
\end{split}
\end{equation}
Since we set $c_{q_2} > c_{q_1}$, $c_d < -1/2$, the form factors $\hat\Pi_1^{2L}$,
$\hat\Pi_1^{d}$ are suppressed compared to $\hat\Pi_1^{1L}$, $\hat\Pi_1^{u}$, and their effect in the
Higgs potential can be neglected.

The fermionic spectrum of the SM fields and heavy resonances of the \mchm\ can be 
expressed in terms of poles and zeros of the form factors
(see Ref.~\cite{Agashe:2005dk} for the gauge spectrum).  
Before EWSB, there are five towers of states:
\begin{itemize}
\item[--] a tower of $q_L$'s (\textbf{2}$_{1/6}$ of SU(2)$_L\times$U(1)$_Y$) 
with masses  given by: zeros$\displaystyle{\left\{\pslash \, \Pi_0^q \right\}}$.
\item[--] a tower of $u_R$'s (\textbf{1}$_{2/3}$ of SU(2)$_L\times$U(1)$_Y$) 
with masses given by: zeros$\displaystyle{\left\{\pslash\, \Pi_0^u \right\}}$.
\item[--] a tower of $d_R$'s (\textbf{1}$_{-1/3}$ of SU(2)$_L\times$U(1)$_Y$) 
with masses given by: zeros$\displaystyle{\left\{\pslash\, \Pi_0^d \right\}}$.
\item[--] a tower of \textbf{2}$_{7/6}$ of SU(2)$_L\times$U(1)$_Y$
with masses given by: poles$\displaystyle{\left\{\pslash
    \left(\Pi_0^u + \frac{1}{2}\, \Pi_1^u \right) \right\}}$.
\item[--] a tower of \textbf{2}$_{-5/6}$ of SU(2)$_L\times$U(1)$_Y$
with masses given by: poles$\displaystyle{\left\{\pslash
    \left(\Pi_0^d + \frac{1}{2}\, \Pi_1^d \right) \right\}}$.
\end{itemize}
After EWSB the different towers are mixed and
the final spectrum consists of:
\begin{itemize}
\item[--] a tower of charge $+2/3$ fermions  with masses given by: \\[0.2cm] \hspace*{0.1cm}
 zeros$\displaystyle{\left\{ p^2 \left(\Pi_0^q+ \frac{\eps^2}{2} \Pi_1^{q_1} \right)
   \left(\Pi_0^u + \frac{\eps^2}{2}\, \Pi_1^u \right) - \frac{\eps^2 (1-\eps^2)}{2} (M_1^u)^2 \right\}}$.
\item[--] a tower of charge $-1/3$ fermions  with masses given by:\\[0.2cm] \hspace*{0.1cm}
 zeros$\displaystyle{\left\{ p^2 \left(\Pi_0^q+ \frac{\eps^2}{2} \Pi_1^{q_2} \right)
   \left(\Pi_0^d + \frac{\eps^2}{2}\, \Pi_1^d \right) - \frac{\eps^2 (1-\eps^2)}{2} (M_1^d)^2 \right\}}$.
\item[--] a tower of charge $+5/3$ fermions
with masses given by: poles$\displaystyle{\left\{\pslash
    \left(\Pi_0^u + \frac{1}{2}\, \Pi_1^u \right) \right\}}$.
\item[--] a tower of charge $-4/3$ fermions 
with masses given by: poles$\displaystyle{\left\{\pslash 
    \left(\Pi_0^d + \frac{1}{2}\, \Pi_1^d \right) \right\}}$.
\end{itemize}
From the formulas above for the fermions of charge $2/3$ and $-1/3$
 one recovers Eq.~(\ref{mf})
by approximating the form factors with their values at $p^2=0$.
Further use of Eqs.~(\ref{finalff5-1}) and (\ref{finalff5-2}) gives
\begin{equation}  \label{muanalytic5}
\begin{split}
m_u\simeq & \, \frac{2}{L_1}\, \eps\, \sqrt{1-\eps^2}\;
 \frac{\;\sqrt{(1/4-c_q^2)(1/4-c_u^2)}\;\M \,(1-\m\M)}
  {\left[(1/2+c_u)(1-\eps^2) + \M^2 \,\big( (1/2+c_q)+ \eps^2 \, \m^2\, (1/2+c_u) \big) \right]^{1/2}} \\[0.15cm]
 & \times 
  \left[ \eps^2 (1/2-c_q) + \M^2 \,\big( 2\, (1/2-c_u) + \m^2 \,(1/2-c_q)(2-\eps^2) \big) \right]^{-1/2}\, ,
\end{split}
\end{equation}
and a similar formula for the down quark mass.

\section{Defining the MCHM$_{10}$}

The quark sector of the \MCHM\ is defined in terms of 5D bulk multiplets
transforming as antisymmetric representations of SO(5).
Each SM generation is identified with the zero modes
of three $\mathbf{10}_{2/3}$ of SO(5)$\times$U(1)$_{X}$,
\begin{equation} \label{fstates}
\begin{aligned}
\xi_q &= \!\left[\!\!
 \begin{array}{ll}
   \mathbf{(2,2)^q_L} = \begin{bmatrix} q'_L(-+) \\ q_L (++)\end{bmatrix} & 
   \mathbf{(2,2)^q_R} = \begin{bmatrix} q'_R(+-) \\ q_R (--)\end{bmatrix} \\[0.45cm]
   \mathbf{(3,1)^q_L}(--)  &  \mathbf{(3,1)^q_R} (++) \\[0.15cm]
   \mathbf{(1,3)^q_L}(--) & \mathbf{(1,3)^q_R}(++)
  \end{array}  \!\!\right] , \\[0.7cm]
\xi_u &=  \!\left[\!\!
 \begin{array}{ll}
   \mathbf{(2,2)^u_L}(+-) & \mathbf{(2,2)^u_R}(-+) \\[0.15cm]
   \mathbf{(3,1)^u_L}(++) & \mathbf{(3,1)^u_R}(--) \\[0.15cm]
   \mathbf{(1,3)^u_L} = 
    \begin{bmatrix} \chi^u_L (++) \\ u^{c\,\prime}_L (-+) \\ d^{c\,\prime}_L (++) \end{bmatrix} &
   \mathbf{(1,3)^u_R} = 
    \begin{bmatrix} \chi^u_R (--) \\ u_R (+-) \\ d'_R (--) \end{bmatrix}
 \end{array} \!\!\right] , \\[0.6cm]
\xi_d &= \!\left[\!\!
 \begin{array}{ll}
   \mathbf{(2,2)^d_L} (+-) & \mathbf{(2,2)^d_R} (-+) \\[0.15cm]
   \mathbf{(3,1)^d_L} (++) & \mathbf{(3,1)^d_R} (--) \\[0.15cm]
   \mathbf{(1,3)^d_L} = 
    \begin{bmatrix} \chi^d_L (++) \\  u^{c\,\prime\prime}_L (++) \\ d^{c\,\prime\prime}_L (-+)\end{bmatrix} &
   \mathbf{(1,3)^d_R} = \begin{bmatrix} \chi^d_R (--) \\ u'_R (--) \\ d_R (+-) \end{bmatrix}
 \end{array} \!\!\right] ,
\end{aligned}
\end{equation}
and an additional $[\widetilde{(\bf{3},\bf{1})}_R \oplus \widetilde{(\bf{1},\bf{3})}_R]$
(an irreducible representation of O(4)) localized on the IR boundary.
Bulk and boundary fields mix through the most general set of O(4)-symmetric 
IR-boundary mass mixing terms:
\begin{equation}
\left[ \mathbf{\overline{(3,1)}_L^{u,d}}  \mathbf{\widetilde{(3,1)}_R} +
       \mathbf{\overline{(1,3)}_L^{u,d}}  \mathbf{\widetilde{(1,3)}_R} \right] + h.c.
\end{equation}
and
\begin{equation} \label{massmixing}
{\widetilde M}_{u,d}
\left[ \mathbf{\overline{(3,1)}_L^{u,d}}  \mathbf{(3,1)^q_R} +
       \mathbf{\overline{(1,3)}_L^{u,d}}  \mathbf{(1,3)^q_R} \right] +
{\widetilde m}_{u,d} \; \mathbf{\overline{(2,2)}_L^q}\, \mathbf{(2,2)^{u,d}_R} + h.c.
\end{equation}
We will denote by $c_i$, $i=q , u , d$, the  bulk masses of each 5D field $\xi_i$
in units of  $k$.
We have grouped the fields of each multiplet $\xi_i$ in
representations of SO(4)$\sim$SU(2)$_L\times$SU(2)$_R$, and used the fact
that an antisymmetric of SO(5) decomposes as 
$\mathbf{10} = \mathbf{4} \oplus \mathbf{6} = \mathbf{(2, 2)} \oplus \mathbf{(1,3)} \oplus \mathbf{(3,1)}$.
A similar 5D embedding also works for the SM leptons, although with different
U(1)$_X$ charges.

The holographic interpretation of the 5D theory defined above is that of three elementary
fields $q_L$, $u_R$, $d_R$ coupled to a CFT sector via the composite operators
$O_q$, $O_u$, $O_d$.~\footnote{We adopt 
a left-source holographic description for $\xi_{q}$ and a 
right-source description for $\xi_{u}$ and $\xi_{d}$.
See Ref.~\cite{Contino:2004vy}. }
This is the same 4D description of the model of Ref.~\cite{Agashe:2004rs}, though
in this case the operators $O_i$ transform as $\mathbf{10}_{2/3}$ representations of the 
SO(5)$\times$U(1)$_X$ global symmetry of the CFT.
Following the usual procedure, we can embed the elementary fields into complete SO(5)$\times$U(1)$_X$
multiplets ($\mathbf{10}_{2/3}$ of SO(5)$\times$U(1)$_X$ in this case),
\begin{equation}
\Psi_{qL} = \begin{bmatrix}
 \begin{pmatrix} q'_L \\ q_L \end{pmatrix} \\[0.4cm]
 (3,1)^q_L \\[0.1cm] (1,3)^q_L \end{bmatrix} \, , \qquad
\Psi_{uR} = \begin{bmatrix}
 (2,2)^u_R \\[0.1cm] (3,1)^u_R \\[0.15cm]
 \begin{pmatrix} \chi^u_R \\ u_R \\ d'_R \end{pmatrix} 
 \end{bmatrix} \, , \qquad
\Psi_{dR} = \begin{bmatrix}
 (2,2)^d_R \\[0.1cm] (3,1)^d_R \\[0.15cm]
 \begin{pmatrix} \chi^d_R \\ u'_R \\ d_R \end{pmatrix} 
 \end{bmatrix} \, 
\end{equation}
and derive the most general holographic Lagrangian at the quadratic order and in momentum space:
\footnote{The operator $\Psi^{ij} \Psi^{kl} \Sigma^m \epsilon^{ijklm}$ 
is also SO(5) invariant, but its form factor identically
vanishes due to the O(4) symmetry of the CFT.
Also, we have omitted for simplicity a possible mixing term between $\Psi_u$ and 
$\Psi_d$, since
it can be safely neglected in our analysis due to the small coupling of $b_R$
to the CFT needed to explain the bottom quark mass.}
\begin{equation}
\begin{split}
{\cal L}_{\rm holo}^{(2)}
 = &\sum_{r=qL,uR,dR}\Big[{\rm Tr}\big(\bar \Psi_r \pslash \, \hat\Pi^{r}_0(p) \Psi_{r}\big)
 +\Sigma\, \bar\Psi_{r}\pslash \, \hat\Pi_{1}^r(p) \Psi_{r}\Sigma^T \Big] \\
  &+\sum_{r=uR,dR}\Big[{\rm Tr}\big(\bar \Psi_{qL} \hat M_0^r(p) \Psi_{r}\big)
 +\Sigma\, \bar \Psi_{qL} \hat M_1^r(p) \Psi_{r}\Sigma^T \Big] + h.c. \, .
\label{effla10}
\end{split}
\end{equation}
The form factors $\hat\Pi$, $\hat M$ can be computed using the holographic technique 
of Ref.~\cite{Agashe:2004rs}:
\begin{equation} \label{finalff10-1}
\begin{alignedat}{2}
\hat\Pi_0^{qL} &= \Pi_{Q_L}(c_q,c_u,\M)\, , & \hat\Pi_1^{qL} &=2\big[\Pi_{q_L}(c_q,c_u,\m)-\Pi_{Q_L}(c_q,c_u,\M)\big] \, ,\\[0.15cm]
\hat\Pi_0^{uR} &=\hat\Pi^{qL}_0(c_q\leftrightarrow c_u,L\leftrightarrow R)\,
,\qquad & \hat\Pi_1^{uR} &=\hat\Pi^{qL}_1(c_q\leftrightarrow c_u,L\leftrightarrow R) \, ,\\[0.15cm]
\hat\Pi_0^{dR} &=\hat\Pi^{qL}_0(c_q\leftrightarrow c_d,L\leftrightarrow R)\,
,\qquad & \hat\Pi_1^{dR} &=\hat\Pi^{qL}_1(c_q\leftrightarrow c_d,L\leftrightarrow R) \, ,\\[0.3cm]
\hat M_0^{uR} &= M_Q(c_q,c_u,\M)\, , & \hat M_1^{uR} &=2\big[M_{q}(c_q,c_u,\m)-M_{Q}(c_q,c_u,\M)\big] \, ,\\[0.15cm]
\hat M_0^{dR} &= M_Q(c_q,c_d,\Md)\, , & \hat M_1^{dR} &=2\sqrt{2}\big[M_{q}(c_q,c_d,\md)-M_{Q}(c_q,c_d,\Md)\big]  \, .
\end{alignedat}
\end{equation}
Here $\Pi_{q_L,Q_L}$ and $M_{q,Q}$ are the form factors defined in the Appendix of 
Ref.~\cite{Agashe:2004rs}. After setting all non-dynamical fields to zero, the Lagrangian
(\ref{effla10}) reduces to that of Eq.~(\ref{effla}) with
\begin{equation} \label{finalff10-2}
\begin{split}
\Pi_0^q &= \hat\Pi^{qL}_0+\frac{1}{2}\, \hat\Pi^{qL}_1 \, , \\
\Pi_0^u &= \hat\Pi^{uR}_0 \, , \\
\Pi_0^d &= \hat\Pi^{dR}_0 \, , 
\end{split} \qquad
\begin{split}
\Pi_1^{q_1}&= -\frac{1}{2}\, \hat\Pi^{qL}_1 \, , \\[0.05cm]
\Pi_1^{q_2}&= -\hat\Pi^{qL}_1 \, , 
\end{split} \qquad 
\begin{split}
\Pi_1^{u}  &= \frac{1}{2}\, \hat\Pi^{uR}_1 \, , \\[0.15cm]
\Pi_1^{d}  &= \frac{1}{2}\, \hat\Pi^{dR}_1 \, , 
\end{split} \qquad
\begin{split}
M_1^u &= \frac{1}{2\sqrt{2}}\, \hat M^{uR}_1  \, , \\[0.05cm]
M_1^d &= \frac{1}{2}\, \hat M^{dR}_1 \, .
\end{split}
\end{equation}

The fermionic spectrum of the \MCHM\ can be 
expressed in terms of poles and zeros of the form factors.
Before EWSB, there are six towers of states:
\begin{itemize}
\item[--] a tower of $q_L$'s (\textbf{2}$_{1/6}$ of SU(2)$_L\times$U(1)$_Y$) 
with masses  given by: zeros$\displaystyle{\left\{\pslash \, \Pi_0^q \right\}}$.
\item[--] a tower of $u_R$'s (\textbf{1}$_{2/3}$ of SU(2)$_L\times$U(1)$_Y$) 
with masses given by: zeros$\displaystyle{\left\{\pslash\, \Pi_0^u \right\}}$.
\item[--] a tower of $d_R$'s (\textbf{1}$_{-1/3}$ of SU(2)$_L\times$U(1)$_Y$) 
with masses given by: zeros$\displaystyle{\left\{\pslash\, \Pi_0^d \right\}}$.
\item[--] a tower of \textbf{2}$_{7/6}$ of SU(2)$_L\times$U(1)$_Y$
with masses given by: poles$\displaystyle{\left\{\pslash\, \Pi_0^q \right\}}$.
\item[--] a tower of \textbf{1}$_{5/3}$ plus a tower of \textbf{3}$_{2/3}$ of 
SU(2)$_L\times$U(1)$_Y$ with masses given by: poles$\displaystyle{\left\{\pslash\, \Pi_0^u \right\}}$.
\end{itemize}
The final spectrum after EWSB consists of:
\begin{itemize}
\item[--] a tower of charge $+2/3$ fermions  with masses given by: \\[0.2cm] \hspace*{0.1cm}
 zeros$\displaystyle{\left\{ p^2 \left(\Pi_0^q+ \frac{\eps^2}{2} \Pi_1^{q_1} \right)
   \left(\Pi_0^u + \frac{\eps^2}{2}\, \Pi_1^u \right) - \frac{\eps^2 (1-\eps^2)}{2} (M_1^u)^2 \right\}}$.
\item[--] a tower of charge $-1/3$ fermions  with masses given by:\\[0.2cm] \hspace*{0.1cm}
 zeros$\displaystyle{\left\{ p^2 \left(\Pi_0^q+ \frac{\eps^2}{2} \Pi_1^{q_2} \right)
   \left(\Pi_0^d + \frac{\eps^2}{2}\, \Pi_1^d \right) - \frac{\eps^2 (1-\eps^2)}{2} (M_1^d)^2 \right\}}$.
\item[--] a tower of charge $+5/3$ fermions 
with masses given by:
 poles$\displaystyle{\left\{\pslash\, \Pi_0^q \right\}}$ \
and \ poles$\displaystyle{\left\{\pslash\, \Pi_0^u \right\}}$.
\end{itemize}
From the formulas above for the fermions of charge $2/3$ and $-1/3$, one recovers Eq.~(\ref{mf})
by approximating the form factors with their values at $p^2=0$.
Further use of Eqs.~(\ref{finalff10-1}) and (\ref{finalff10-2}) 
gives the same explicit formula valid for the \mchm, 
Eq.~(\ref{muanalytic5}), but a factor $\sqrt{2}$ smaller:
\begin{equation} \label{muanalytic10}
m_u|_{\text{MCHM}_{10}} \simeq \frac{1}{\sqrt{2}}\, m_u|_{\text{MCHM}_5} \, .
\end{equation}
A similar result is also valid for the down quark mass.

%%%%%%%%%%
%%%%%%%%%%    References
%%%%%%%%%%

\end{document}